\documentclass[epj]{svjour}
\usepackage{graphics}

\usepackage{lineno,hyperref}
\modulolinenumbers[5]
\usepackage{graphicx}
\usepackage{subfig}
\usepackage{caption}
\usepackage[utf8]{inputenc}
\usepackage{graphicx}
\usepackage{amsmath}
\usepackage{arydshln}
\usepackage{lineno,hyperref}
\modulolinenumbers[5]
\usepackage{graphicx}
\usepackage{subfig}
\usepackage{caption}
\usepackage{color}
\usepackage[utf8]{inputenc}
\usepackage{amsfonts}
\usepackage{graphicx,amssymb,amsmath}
\usepackage{amscd}
\usepackage[english]{babel}
\usepackage{algorithmic}
\usepackage{algorithm}
\usepackage{multirow}
\begin{document}
\title{Study of $p_{T}$ spectra of light particles using modified Hagedorn function and cosmic rays Monte Carlo event generators in proton-proton collisions at  $\sqrt s$  = 900 GeV}

\author{Muhammad Ajaz\inst{1}, Muhammad Waqas\inst{2}, Guang Xiong Peng\inst{2}, Zafar Yasin\inst{3},  Hannan Younis\inst{4} \and Abd Al Karim Haj Ismail\inst{5,6}
}                     
%
%
\institute{Department of Physics, Abdul Wali Khan University Mardan, 23200 Mardan, Pakistan \and School of Nuclear Science and Technology, University of Chinese Academy of Sciences, Beijing 100049, China
\and Pakistan Institute of Nuclear Science and Technology (PINSTECH) Islamabad, 44000 Islamabad, Pakistan
\and Department of Physics, COMSATS  University Islamabad, 44000 Islamabad, Pakistan
\and College of Humanities and Sciences, Ajman University, 346 Ajman UAE \and Nonlinear Dynamics Research Center (NDRC), Ajman University, 346 Ajman UAE}
\date{Received: date / Revised version: date}
%
\abstract{
Transverse momentum spectra ($p_T$) of charged particles including $\pi^{\pm}$, $K^{\pm}$ and (anti-)protons measured by ALICE experiment in the $p_T$ range of 0.1~--~2.5 GeV/$c$ and $|\eta|$ $<$ 0.5 are studied in $pp$ collisions at $\sqrt s$ = 900 GeV using modified Hagedorn function with embedded transverse flow velocity and are compared to the predictions of EPOS--LHC, Pythia, QGSJET and Sibyll models. We find that the average transverse flow velocity ($\beta_T$) decreases with increasing the mass of the particle while the kinetic freeze-out temperature ($T_{0}$) extracted from the function increases with the particle’s mass. The former varies from (0.36 $\pm$ 0.01) c to (0.25 $\pm$ 0.01) c for $\pi^{\pm}$ to protons while the latter from (76 $\pm$ 6) MeV to (95 $\pm$ 5) MeV respectively. The fit of the models predictions also yield the same values for $T_{0}$ and $\beta_T$ as the experimental data. The only difference is in the values of $n$, and $N_0$ which yields different values for different models. The EPOS--LHC, Pythia, and QGSJET models reproduce the data in most of the $p_{T}$ range for $\pi^{\pm}$, EPOS--LHC and Sibyll for $K^{\pm}$ up to 1.5 GeV/$c$ and EPOS--LHC for protons up to 1.6 GeV/$c$. The model simulations also reproduced the behavior of increasing average transverse momentum with mass reported by the ALICE experiment. 
\PACS{
      {13.85.−t}{Hadron-induced high- and super-high-energy interactions},
      {25.75.Nq} \and
      {13.85.Hd}{discribing text of that key}
     } 
} 
\maketitle
\section{Introduction}
\label{intro}
The study of hadron production is being used in cosmic ray physics as well as in nuclear and particle physics by absolute yield, angular distributions, transverse momentum ($p_{T}$), rapidity ($y$) and pseudo-rapidity ($\eta$) distributions. These studies can be used to tune the event generators including but not limited to partons interaction, partons correlation, partons hadronization, collective flow, spin, and final state effects etc \cite{1}. Furthermore, centrality dependence of the yield and hadrons spectra can be used to improve the related parameters in the modeling of the event generators. \\
The study of $pp$ collisions is further relevant to the investigation of heavy-ion collisions as it is being used as a reference to find the properties of medium effect. In case of heavy-ion collisions, various final-state effects are modifying the yield as well as the spectral shape of different hadron \cite{2,3}.\\
A Tsallis distribution function has been used by many researchers because of its very good reproduction of the $p_{T}$ spectra of experimental data at low as well as high $p_{T}$. The Tsallis distribution function is further extended by including different transverse flow models to describe the $p_{T}$ spectra of hadrons produced in $pp$ and heavy-ion collisions at high-energies. In literature, the blast-wave model coupling with Boltzmann–Gibbs statistics (BGBW model) \cite{4,5}, the blast-wave model with Tsallis statistics (TBW model) \cite{6}, the Tsallis distribution incorporating the transverse flow effect also known as Improved Tsallis distribution as well as incorporating other model functions \cite{6,7}, and Hagedorn function with transverse flow \cite{8,9,10,11} have been used to extract the kinetic freeze-out temperature and transverse flow velocity. 
 In continuation of \cite{11a,11b,11c,11d,11e,11f,11g}, here we used the measurements of the charged pions and kaons, and (anti-)protons in $pp$ collision by the ALICE collaboration at 900 GeV \cite{12}.  Simulation results of four different hadron production models are used for comparison with the data with the same cuts used by the data. Furthermore, we have used the Hagedorn function with the embedded transverse flow velocity to extract the kinetic freeze-out temperature and transverse flow velocity.\\

\section{The method and models}
\label{method}
Transverse momentum spectra ($p_T$) of $\pi^{\pm}$, $K^{\pm}$, and (anti-) protons in $pp$ collisions at $\sqrt s$ = 900 GeV measured by the ALICE experiment \cite{12} in the transverse momentum range of $p_{T}$ = 0.1~--~2.5 GeV/$c$ and pseudorapidity region of $|\eta|$ $<$ 0.5 are investigated by using modified Hagedorn function with embedded transverse flow velocity. The measured spectra are also compared to the predictions of Cosmic Ray Monte Carlo (crmc) models including EPOS--LHC \cite{13}, Pythia \cite{14}, QGSJET \cite{15,15a}, and Sibyll2.3d \cite{16}. Simulations of one million $pp$ collision at 900 GeV under the same condition as the data were performed by using the above event generators. \\
Several hadron production models are being used to simulate air showers including the ones used in this study. The advantage of these models is that that they are tuned to LHC data and can be used up to 7 TeV energies for reproducing results of the experimental data. All these event generators are mainly based on simple Parton models incorporating Gribov-Regge theory. Multiple scattering approach is used in these models that are resulting exchange of multiple Parton ladders between target and projectile. Beside this, every model has a different philosophy. 
The EPOS model is used both for hadron-hadron, nucleus-nucleus interaction and simulation of cosmic rays air showers. It is a minimum bias hadronic interaction model aiming to describe soft particle production below 5 GeV/$c$ with greater details at any energy and mass of the colliding system. Several effects pertaining to heavy-ion collisions are incorporated in it such as screening \cite{17}, Cronin effect, collective effects \cite{18}, and Parton saturation. The same covariant approach is used as the previous version of the model with an improved flow parameterization. New features included in the model are described in greater detail in Ref. \cite{13}.
The Pythia Monte Carlo event generator is developed to generate interactions of high-energy particles to produce their properties in strong interaction. The code is mainly based on a different QCD-based models along with the use of some analytical results. The model incorporates processes such as hard sub processes, final- and initial-state Parton showers, underlying events, beam remnants, decays and fragmentation. A detail description of the model can be found in Ref. \cite{14}.
QGSJET model, based on Quark Gluon Sting (QGS) \cite{19} model using Gribov-Regge theory, is a model commonly used for cosmic ray simulations by different group of researchers and collaborations \cite{15} for many years. \\
At high energies, the transverse momentum at the high $p_{T}$ tail of the hadrons spectra in hadron-hadron and nucleus-nucleus collisions is very well reproduced by QCD inspired Hagedorn function \cite{8}:\\
\begin{align}
\label{eqa}
\frac{d^2N}{2\pi N_{ev}p_{T}dp_{T}dy} = C\bigg(1+\frac{m_T}{p_0}\bigg)^{-n},
\end{align}
In this expression, $m_{t}$ = $\sqrt{p_T^2 + m_0^2}$, is the transverse mass with $m_0$ as the rest mass of a hadron, $C$ is the constant of the fit, and $n$ and $p_{0}$ are the two free parameters of the function to be determined. The term $p_{0}$ in Eq. $\ref{eqa}$ is equal to $nT_{0}$, where $T_{0}$ is used to incorporate the temperature (kinetic freeze-out temperature) in the function. The equation becomes
\begin{align}
\label{eqb}
\frac{d^2N}{2\pi N_{ev}p_{T}dp_{T}dy} = C\bigg(1+\frac{m_T}{nT_0}\bigg)^{-n},
\end{align}
To incorporate the value of the collective transverse flow in Eq. \ref{eqb}, $m_{T}$ is replaced with
\begin{align}
\label{eqc}
m_{T} = <\gamma_{T}>\bigg(m_{T}-p_{T}<\beta_{T}>\bigg)
\end{align}
The last modification has been used by \cite{9,10,16a}, the final equation becomes

\begin{align}
\label{eqd}
\frac{d^2N}{2\pi N_{ev}p_{T}dp_{T}dy} =C \bigg(1+<\gamma_{T}>\frac{m_{T}-p_{T}<\beta_{T}>}{nT_0}\bigg)^{-n},
\end{align}
Where $C$ is the normalization constant which results in the integral of Eq. \ref{eqd} to be normalized to 1, $T_{0}$ is the kinetic freeze-out temperature, $<\gamma_{t}>$ = $1/\sqrt{1-<\beta_{t}^{2}>}$, and $<\beta_t>$ is the average transverse flow velocity. 
The last equation is known as Hagedorn function with the embedded transverse flow velocity. It is worth mentioning that the function has been used in Refs. \cite{9,10,16a}. With only a few parameters, the model has successfully described the physical parameters of the spectra as described in \cite{9,10,16a}.
The current form of the modified Hagedorn function is further modified for the current analysis as follows:
\begin{align}
\label{eqe}
\frac{d^2N}{N_{ev}dp_{T}dy} = 2\pi p_{T} C \bigg(1+<\gamma_{T}>\frac{m_{T}-p_{T}<\beta_{T}>}{nT_0}\bigg)^{-n},
\end{align}
We used the above Hagedorn formula (Eq. \ref{eqe}) to fit the $p_{T}$ spectra of charged pions, kaons, and (anti-)protons and extracted the parameters for the data as well as all the models for comparison. \\

\subsection{Results and Discussion}
\label{results}
\begin{figure}
\centering
\includegraphics[width=0.40\textwidth]{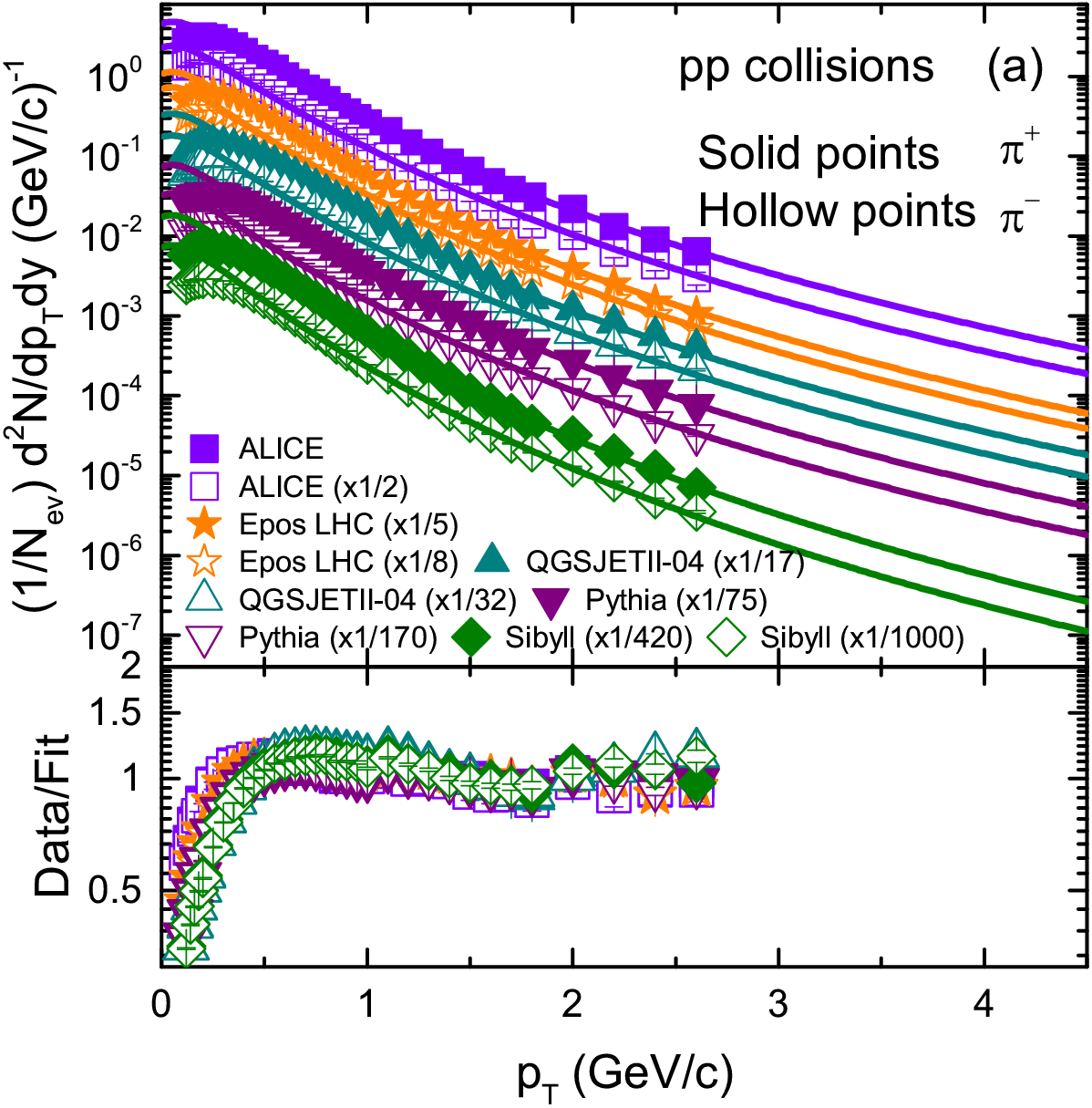}
\includegraphics[width=0.40\textwidth]{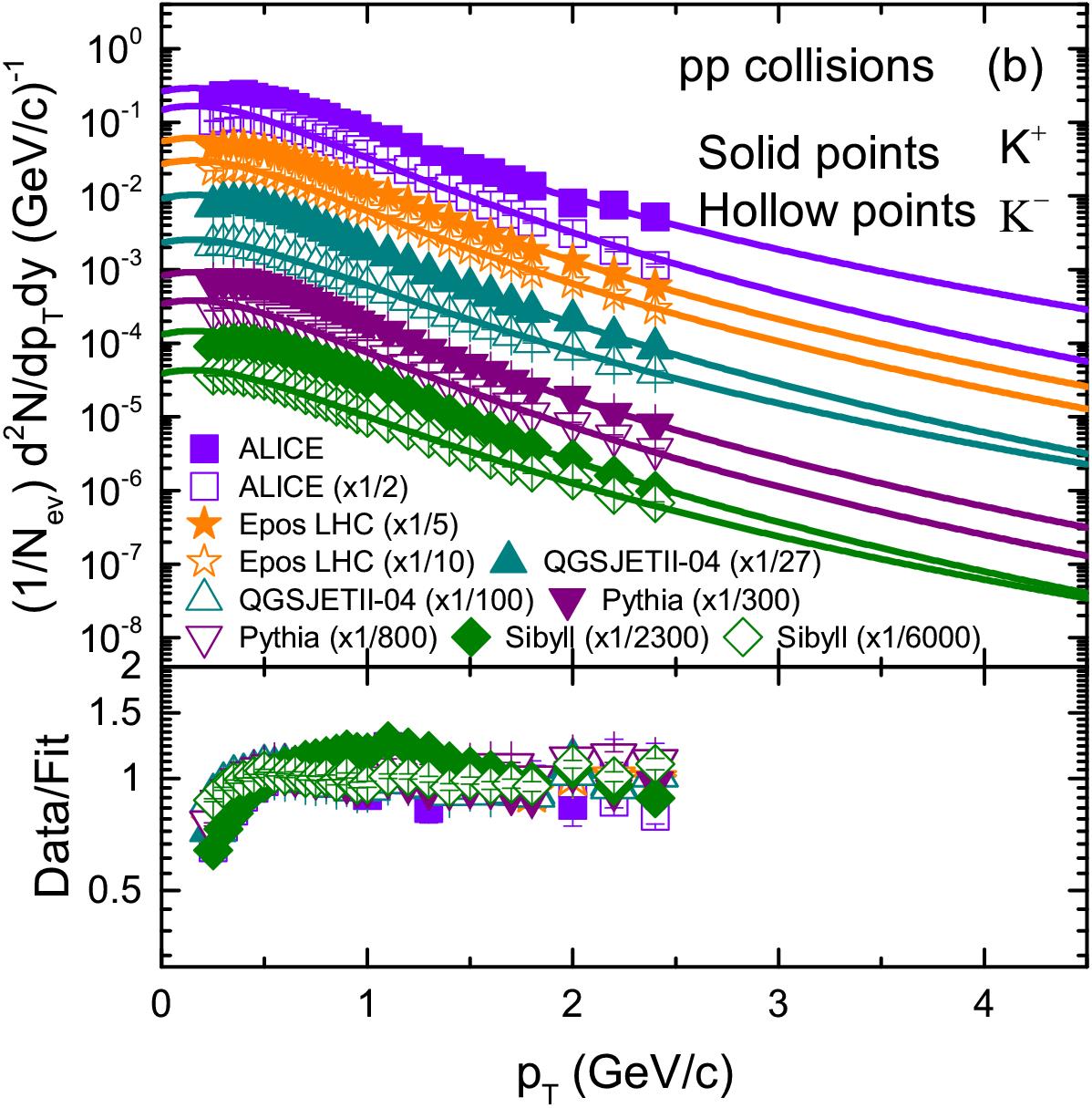}
\includegraphics[width=0.40\textwidth]{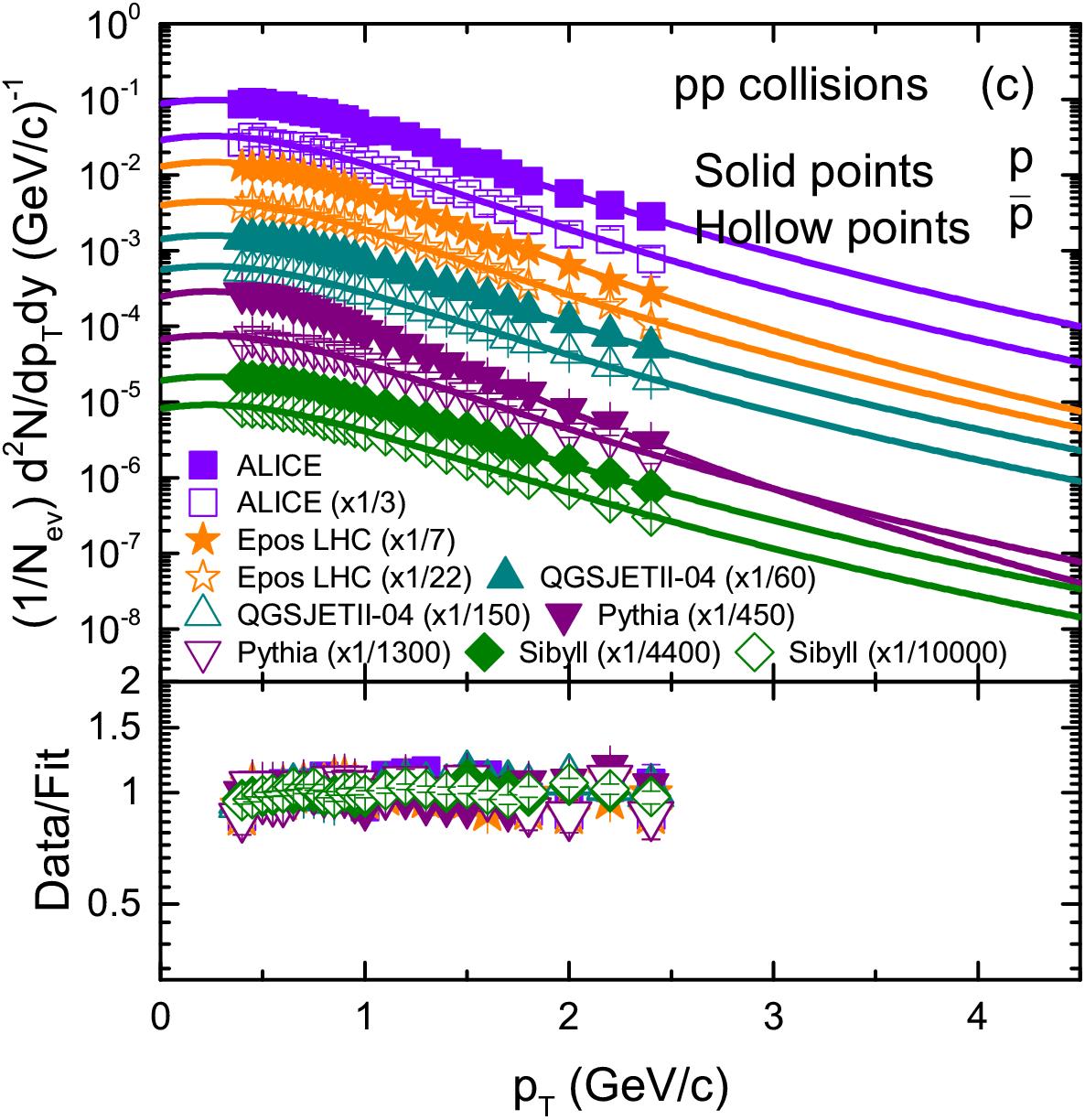}\\ \vspace{.35cm}

\caption{Result of the fit function on the transverse momentum spectra of pions (a) kaons (b) and protons (c) produced in $pp$ collision at 900 GeV by Data, EPOS--LHC, Pythia, QGSJET and Sibyll models. Filled markers are used to represent positive particles while open markers are used to show negative particles. Lines in different colors are used to show the fit curve obtained using Eq \ref{eqe}.}
\label{fig:1}
\end{figure}

The transverse momentum spectra ($p_T$) of identified charged particles measured by the ALICE experiment in the $p_{T}$ range of (0.1--2.5) GeV/$c$ and pseudorapidity $|\eta|$ $<$ 0.5 in $pp$ collisions at $\sqrt s$ = 900 GeV are analysed by using a modified Hagedorn function with the embedded transverse flow velocity (Eq. \ref{eqe}). The results of the fit function are given in Table 1. The result of the fit curves on the $p_T$ distributions of the particles are shown in Figures \ref{fig:1}. The solid and hollow symbols in each panel represent the data for particles and their anti-particles respectively.
Figure 1(a). shows the fit curves for the experimental data and models for $\pi^{\pm}$ mesons, while figure 1(b) and 1(c) show the same for $K^{\pm}$ and (anti)-protons respectively. The curve shows that the function fit the experimental data approximately well. In case of the pions, there is some discrepancy between the data/fit and fit function result for $p_T$ $<$ 0.5 GeV/$c$, which is indicated by the Data/Fit curve shown in the lower panel of the plot. The reason behind this discrepancy is that most of the low energy pions are produced by the decay of resonance particles -- an effect which is not incorporated into the function. The Data/Fit above $p_T$ = 0.5 GeV/$c$ is around unity in Fig. 1(a) as well as in Fig 1(b) and 1(c) which shows good agreement of the fit function to the experimental data.

\begin{table}
\label{tab:1}
{\scriptsize Table 1. The values of free parameters
($n$, $T_0$ and $<\beta_T>$) and normalization constant ($N_0$) obtained by the fit function (Eq. \ref{eqe}) from the $p_T$ distribution of charged particles shown in figure 1.
\begin{center}
\begin{tabular}{cccccccccc}\\ \hline\hline
Collisions         &   Data      & Particle         & $T_0$           & $\beta_T$         &    $n$       &    $N_0$      \\ \hline
Fig. 1(a)             &   ALICE     & $\pi^+$          &$0.076\pm0.006$  & $0.360\pm0.010$  & $7.4\pm0.5$  & $196\pm12$ \\
                   &   ALICE     & $\pi^-$          &$0.076\pm0.005$  & $0.360\pm0.009$  & $7.4\pm0.5$ & $196\pm11.5$ \\
   $pp$              &   Epos-LHC  & $\pi^+$          &$0.076\pm0.005$  & $0.360\pm0.010$  & $7.6\pm0.4$  & $220\pm21$ \\
                   &   Epos-LHC  & $\pi^-$          &$0.076\pm0.006$  & $0.340\pm0.010$ & $7.6\pm0.8$  & $210\pm25$ \\
   0.9 TeV         &   QGSJET11  & $\pi^+$          &$0.076\pm0.005$  & $0.360\pm0.008$  & $7.6\pm0.6$  & $226\pm25$ \\
                   &   QGSJET11  & $\pi^-$          &$0.076\pm0.005$  & $0.360\pm0.009$  & $7.6\pm0.5$  & $226\pm19$ \\
                   &   Pythia    & $\pi^+$          &$0.076\pm0.004$  & $0.360\pm0.009$  & $7.6\pm0.6$  & $220\pm20$ \\
                   &   Pythia    & $\pi^-$          &$0.076\pm0.005$  & $0.360\pm0.008$  & $7.6\pm0.7$  & $215\pm23$ \\
                   &   Sibyll    & $\pi^+$          &$0.076\pm0.006$  & $0.360\pm0.008$  & $8.6\pm0.6$  & $270\pm31$ \\
                   &   Sibyll    & $\pi^-$          &$0.076\pm0.005$  & $0.360\pm0.008$  & $8.6\pm0.5$  & $270\pm19$ \\
\hline
Fig. 1 (b)            &  ALICE       & $K^+$          &$0.086\pm0.004$  & $0.300\pm0.011$  & $6.5\pm0.6$ & $23\pm2$ \\
                   &  ALICE       & $K^-$          &$0.086\pm0.005$  & $0.300\pm0.011$  & $7.5\pm0.5$ & $24\pm3$ \\
   $pp$              &  Epos-LHC    & $K^+$          &$0.086\pm0.006$  & $0.300\pm0.011$  & $7.3\pm0.6$ & $22.5\pm2$ \\
                   &  Epos-LHC    & $K^-$          &$0.086\pm0.006$  & $0.300\pm0.010$  & $7.3\pm0.4$  & $22.3\pm2$ \\
   0.9 TeV         &  QGSJET11    & $K^+$          &$0.086\pm0.005$  & $0.300\pm0.009$  & $7.6\pm0.3$  & $20\pm2$ \\
                   &  QGSJET11    & $K^-$          &$0.086\pm0.005$  & $0.300\pm0.009$  & $6.6\pm0.3$  & $20\pm1$ \\
                   &  Pythia      & $K^+$          &$0.086\pm0.005$  & $0.300\pm0.010$  & $7.5\pm0.4$  & $20\pm3$ \\
                   &  Pythia      & $K^-$          &$0.086\pm0.005$  & $0.300\pm0.010$  & $7.5\pm0.4$  & $22\pm3$ \\
                   &  Sibyll      & $K^+$          &$0.086\pm0.004$  & $0.300\pm0.010$  & $7.7\pm0.7$  & $24\pm3$ \\
                   &  Sibyll      & $K^-$          &$0.086\pm0.004$  & $0.300\pm0.010$  & $8.6\pm0.7$  & $20\pm4$ \\
\hline
Fig. 1(c)             &  ALICE       & $p$          &$0.095\pm0.005$  & $0.250\pm0.010$  & $8\pm0.5$ & $10\pm0.6$ \\
                   &  ALICE       & $\bar p$     &$0.095\pm0.005$  & $0.250\pm0.010$  & $8\pm0.5$ & $10\pm0.6$ \\
   $pp$              &  Epos-LHC    & $p$          &$0.095\pm0.006$  & $0.250\pm0.010$  & $8.8\pm0.6$& $10\pm0.4$ \\
                   &  Epos-LHC    & $\bar p$     &$0.095\pm0.005$  & $0.250\pm0.010$  & $8\pm0.5$  & $10\pm0.3$ \\
   0.9 TeV         &  QGSJET11    & $p$          &$0.095\pm0.005$  & $0.250\pm0.009$  & $7.6\pm0.3$  & $10\pm2$ \\
                   &  QGSJET11    & $\bar p$     &$0.095\pm0.005$  & $0.250\pm0.009$  & $7.6\pm0.3$  & $9.8\pm1$ \\
                   &  Pythia      & $p$          &$0.095\pm0.004$  & $0.250\pm0.011$  & $10.3\pm0.7$  & $11.6\pm1.4$ \\
                   &  Pythia      & $\bar p$     &$0.095\pm0.004$  & $0.250\pm0.010$  & $8\pm0.6$    & $10\pm1.2$ \\
                   &  Sibyll      & $p$          &$0.095\pm0.005$  & $0.250\pm0.010$  & $7.5\pm0.4$  & $10\pm1$ \\
                   &  Sibyll      & $\bar p$     &$0.095\pm0.004$  & $0.250\pm0.010$  & $7.5\pm0.4$  & $9.7\pm1.1$ \\
\hline
\end{tabular}%
\end{center}}
\end{table}

 Table 1. shows the values of the free parameters obtained by the fit function in Eq. \ref{eqe}. The average transverse flow velocity ($<\beta_T>$) decreases with increasing the mass of the particle (which is a natural hydro-dynamical behavior). It is equal to (0.36 $\pm$ 0.01)c for $\pi^{\pm}$, (0.03 $\pm$ 0.01)c for $K^{\pm}$ mesons while (0.25$\pm$0.01)c for $p~(\bar p)$. Similarly the value $T_0$ increases with increasing the mass of the particles with (76$\pm$6) MeV, (86$\pm$5) and (95$\pm$5) MeV respectively for $\pi^{\pm}$, $K^{\pm}$ and $p~(\bar p)$ which shows the mass dependent kinetic freeze-out scenario, in agreement with ref. \cite{20,21,22,23}. The fit of the function on the models predictions also yield the same values for $T_{0}$ and $\beta_T$ as the experimental data. The difference between the models and data is only reflected in the difference of $n$ and $N_0$ which also have comparable values. The parameter $N_0$ is observed to be dependent on the mass of the particle. The parameter $n$ is connected to the scattering centers involved in the interaction process and yield a higher value for multiple scattering center, while $N_0$ is the fit constant which is used for the comparison of the fit function.
 It should be noted that the normalization constants $C$ in Eq. \ref{eqe} and $N_0$ in table 1 are not the same. $C$ is the normalization constant which is used to let the integral of Eq. \ref{eqd} be 1, while the normalization constant $N_0$ is used to compare the fit function and the experimental spectra. The constant $C$ can be absorbed in $N_0$, but we have used both $C$ and $N_0$ to give a clear description. The parameter $N_0$ has its physical significance. It reflects the multiplicity. In the present work, we found that $N_0$ is larger for pions, followed by kaons, while protons have the lowest value for $N_0$ which means that pions have larger multiplicity and protons have the lowest multiplicity. This result is in agreement with \cite{24}.

\begin{figure}
\centering
\includegraphics[width=0.40\textwidth]{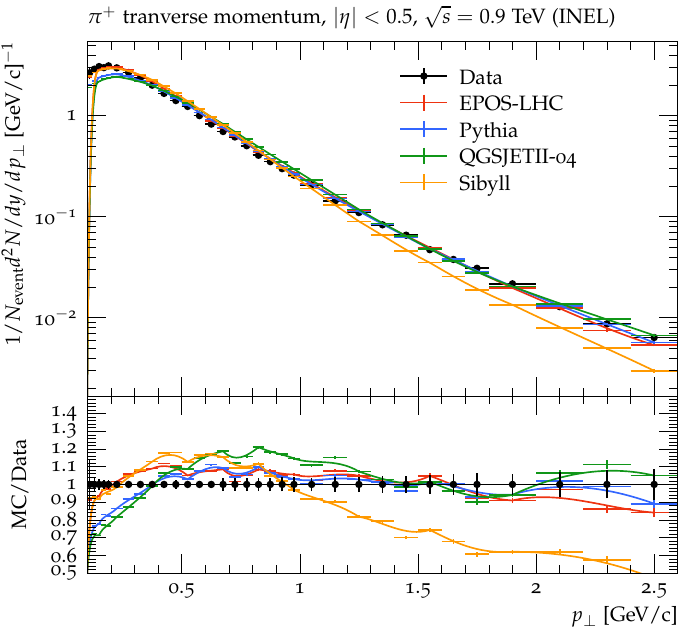}
\includegraphics[width=0.40\textwidth]{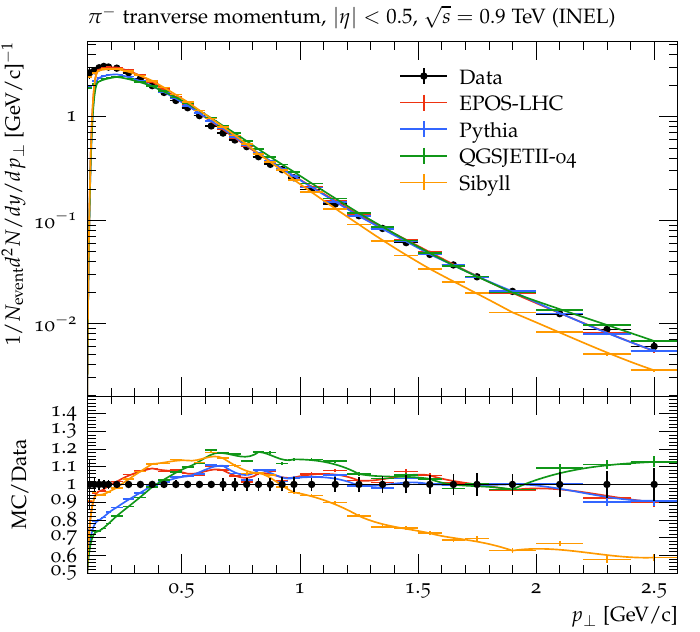}\\ \vspace{.35cm}
\includegraphics[width=0.40\textwidth]{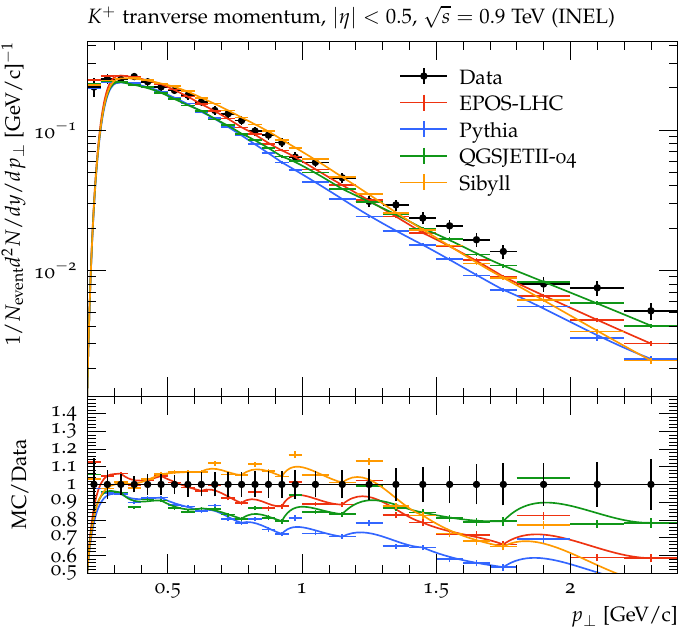}
\includegraphics[width=0.40\textwidth]{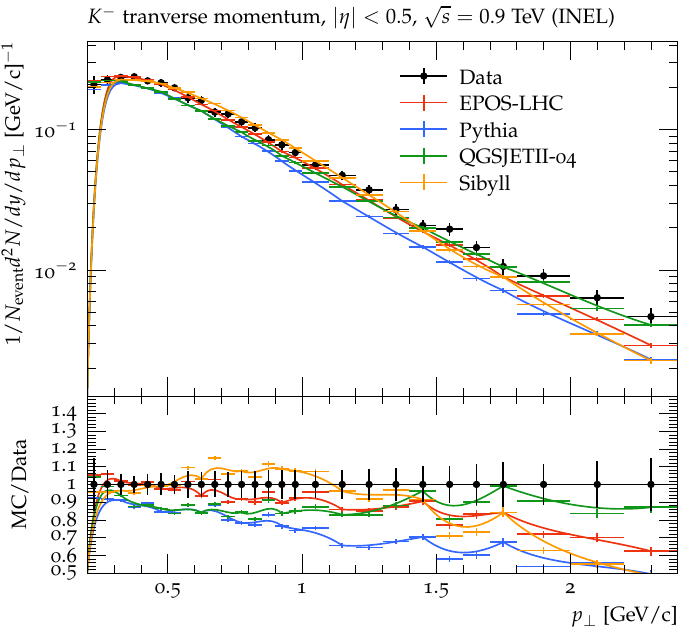}\\ \vspace{.35cm}
\includegraphics[width=0.40\textwidth]{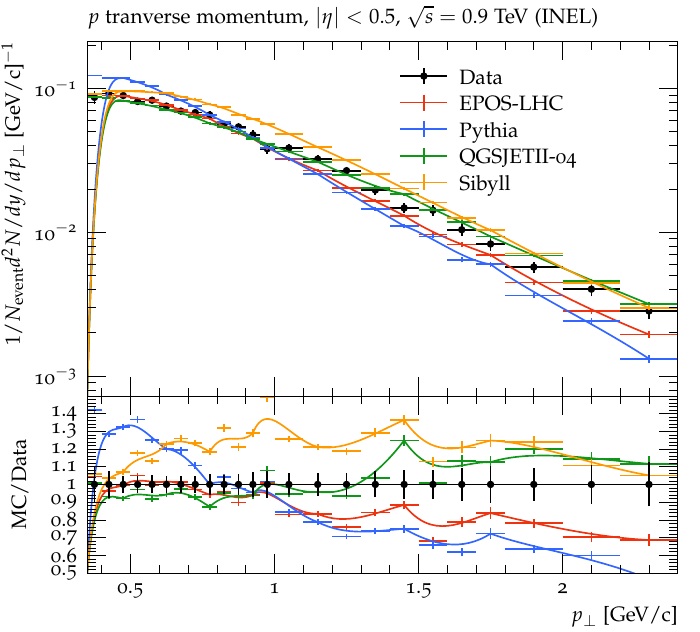}
\includegraphics[width=0.40\textwidth]{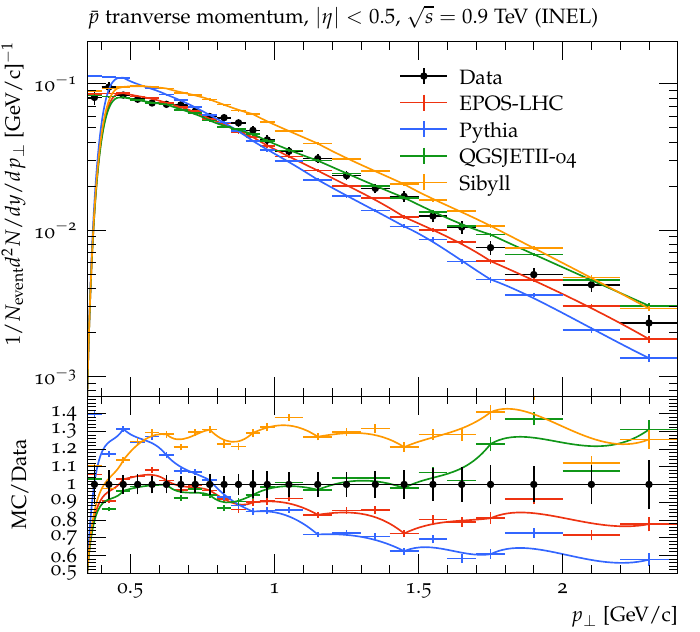}\\ \vspace{.35cm}

\caption{Transverse momentum spectra of the positively charged (left column) and negatively charged (right column) pions (first row) kaons (second row) and protons (third row) produced in $pp$ collision at 900 GeV by EPOS--LHC, Pythia, QGSJET and Sibyll models are compared with the measurement of ALICE experiment. Filled black circles are used to represent experimental data while lines of different colors are used to show the models results as shown in the legend. The lower panel of the graph shows the ratio of the model to data.}
\label{fig:4}
\end{figure}

Transverse momentum spectra of the positively charged (left column) and negatively charged (right column) pions (first row) kaons (second row) and protons (third row) produced in $pp$ collision at 900 GeV by EPOS--LHC, Pythia, QGSJET and Sibyll models are compared with the measurement of ALICE experiment. The solid black markers represents experimental data while line of different colors are used to show the models result highlighted in the legend of the graph. In case of experimental data, the vertical error bars are the quadrature addition of statistical and systematic errors, while in case of models predictions, the error bars represents statistical errors. Furthermore, the horizontal bar at a point is the width of the x-axis bin. The lower panel of the graph shows the ratio of the model to data.
As can be seen from Figure \ref{fig:1}, all the models with in the error bars can reproduce the $\pi^{\pm}$ spectra very well over the entire $p_T$ range except Sibyll which underestimates the data above 1 GeV/$c$ by up to 40 $\%$. In the low $p_T$ range of less than 0.2 GeV/$c$, Pythia and QGSJET underestimate the data by up to 25 $\%$. For $K^{\pm}$, the EPOS--LHC and QGSJET models reproduce the data up to 1.3 GeV/$c$ with QGSJET describing the data for $K^-$ meson over the entire $p_T$ range. The other two models underestimate the data having increasing discrepancy with increasing $p_T$. The QGSJET model reproduce the protons and anti-protons spectra up to 1.4 GeV/$c$ and 1.7 GeV/$c$ respectively while the EPOS--LHC model could describe the data only at low $p_T$ up to 0.9 GeV/$c$. Pythia and EPOS--LHC underestimate the data above 1 GeV/$c$ while Sibyll overestimate the distribution over the entire $p_T$ range.

\begin{figure}
\centering
\includegraphics[width=0.80\textwidth]{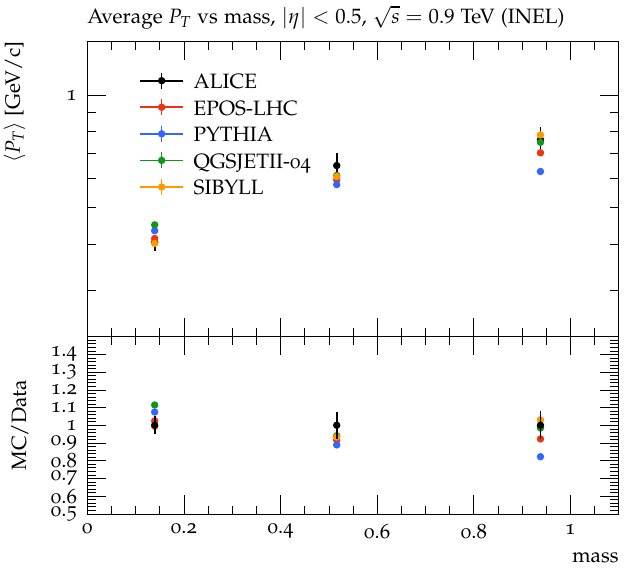}
\caption{The average transverse momentum as a function of mass for the particles in the current study are shown for experimental data by filled black circles while filled circular markers of different colors are used for models predictions. The lower panel of the graph shows the ratio of the models to the experimental data.}
\label{fig:5}
\end{figure}

Fig. \ref{fig:5} shows the variation of the average transverse momentum $<p_T>$ as a function of the mass of the particles. The $<p_T>$ is observed to be increasing with increasing mass of the particles. The behavior of increasing $<p_T>$ with increasing the mass of the particle is well reproduced by all the models. Only the Pythia model underestimate the $<p_T>$ for protons only.

\subsection{Summary and Conclusions}
\label{conclusion}
Analyses of the transverse momentum spectra of identified charged particles are presented using modified Hagedorn function with embedded transverse flow velocity. The function fit the data approximately well indicated by the Data/Fit curves. We found that the average transverse flow velocity ($\beta$) decreases while the kinetic freezeout temperature increases with increasing the mass of the particle. The same results has been reproduced by the models as well. For $p_T$ $<$ 0.5 GeV/$c$, the function could not fit the data and model simulations very well. This is the region where most of the pions are produced by the decay of resonances -- an effect that has not been considered in the function. for region with  $p_T$ $>$ 0.5 GeV/$c$, the function fit all the data very well. 
Simultaneously,  the measurements performed by the ALICE experiment were compared by prediction of several event generator presented in the transverse momentum range of $p_{T}$ = 0.1--2.5 GeV/$c$ and pseudorapidity $|\eta|$ $<$ 0.5 in $pp$ collisions at $\sqrt s$ = 900 GeV. 
Except Sibyll which underestimates the data above 1 GeV/$c$ by up to 40 $\%$, all models with in the error bars reproduced the pions spectra well over the entire $p_T$ range. The EPOS--LHC and QGSJET models reproduced the kaons spectra up to 1.3 GeV/$c$ with QGSJET describing the data for $K^-$ meson over the entire $p_T$ range. The other two models underestimate the data with a larger discrepancy with increasing $p_T$. The QGSJET model reproduce the $p~(\bar p)$ spectra up to 1.4 GeV/$c$ and 1.7 GeV/$c$ respectively while the EPOS--LHC model could describe the data only at low $p_T$ up to 0.9 GeV/$c$. The average transverse momenta of the particles increases with increasing the mass of the particle. The behavior of increasing average transverse momentum with increasing the mass of the particle is well reproduced by all the models. Only the Pythia model underestimated the average transverse momentum for protons only. Although all the model reproduced the measurements by the ALICE experiment for some of the particles or for a region of the momentum range, none of them reproduced the spectra for all the particles over the entire $p_T$ range.

%
%

%
%

\end{document}